\documentclass[
  a4paper,
  11pt,
  aps,prd,
  showkeys,
  nofootinbib,
  ]{revtex4}

\usepackage[utf8x]{inputenc}
\usepackage[T1]{fontenc}
\usepackage{amsmath}
\usepackage{amsfonts}
\usepackage{amssymb}
\usepackage[amssymb]{SIunits}
\usepackage{bm}
\usepackage{braket}
\usepackage{cancel}
\usepackage{cool}
\usepackage{graphicx}
\usepackage[
  pdftex,
  colorlinks=true,
  bookmarksopen=true,
  bookmarksopenlevel=1,
  ]{hyperref}
\usepackage[english]{babel}
\usepackage[babel=true]{microtype}
\usepackage{lmodern}
\usepackage{booktabs}

\newcommand{\sstop}{\tilde{t}}

\newcommand{\neutralino}{\tilde{\chi}^0}
\newcommand{\chargino}{\tilde{\chi}^\pm}

\newcommand{\Kaimini}{\texttt{Kaimini}}
\newcommand{\Minuit}{\texttt{Minuit}}
\newcommand{\Migrad}{\texttt{Migrad}}
\newcommand{\SPheno}{\texttt{SPheno}}

\newcommand{\sol}{\mathrm{sol}}
\newcommand{\atm}{\mathrm{atm}}

\newcommand{\RpV}{$R$-parity breaking}

\newcommand{\BR}[2]{\mathrm{BR}(#1 \to #2)}
\newcommand{\chisqmin}{\chi^2_{\mathrm{min}}}

\newcommand{\Spot}{\mathcal{W}}
\newcommand{\signcond}{\epsilon_2 \epsilon_3 / (\Lambda_2 \Lambda_3)}

\newcommand{\Let}{\mathrel{\mathop:\!\!=}}

\def\gsim{\raise0.3ex\hbox{$\;>$\kern-0.75em\raise-1.1ex\hbox{$\sim\;$}}}
\def\lsim{\raise0.3ex\hbox{$\;<$\kern-0.75em\raise-1.1ex\hbox{$\sim\;$}}}

\begin{document}

\title{Determining $R$-parity violating parameters from neutrino
and LHC data}

\author{F. Thomas}
\email{fthomas@physik.uni-wuerzburg.de}

\author{W. Porod}
\email{porod@physik.uni-wuerzburg.de}

\affiliation{Institut f\"ur Theoretische Physik und Astrophysik,
  Universit\"at W\"urzburg, Am Hubland, 97074 W\"urzburg, Germany}

\begin{abstract}
In supersymmetric models neutrino data can be explained by $R$-parity
violating operators which violate  lepton number by one
unit. The so called bilinear model can account for the observed neutrino
data and predicts at the same time several decay properties of the
lightest supersymmetric particle. In this paper we discuss the expected
precision to determine these parameters by combining neutrino and LHC
data and discuss the most important observables. We show that one
can expect a rather accurate determination of the underlying $R$-parity
parameters assuming mSUGRA relations between the $R$-parity conserving ones
and discuss briefly also the general MSSM as well as the expected accuracies
in case of a prospective $e^+ e^-$ linear collider.
An important observation is that several parameters can only be determined
up to relative signs or more generally relative phases.
\end{abstract}


\maketitle

\section{Introduction}
\label{sec:introduction}

With the start of the LHC the exploration of the terascale has begun
and, thus, in the search for extensions of the Standard Model (SM)
significant higher mass scales can be tested as in the past.
Supersymmetry (SUSY) is  among the most favoured candidates
as it allows for example the unification of gauge couplings and
stabilizes the hierarchy of the Planck scale
(or the GUT scale) and the electroweak scale. Once supersymmetry has
been found, there will be two main tasks: (i) to determine the underlying
parameters of the model and (ii) to check if the minimal model is realized
in nature or an extended one.

In principle one can supersymmetrize all known mechanisms to
generate neutrino masses and mixing angles, see e.\,g.\
ref.~\cite{Nath:2010zj}. However, supersymmetry
offers an intrinsic possibility to generate neutrino masses,
namely $R$-parity violation (RPV), for an review see e.\,g.\
\cite{Hirsch:2004he}.
In this case neutrino masses are generated either via
mixing with neutralinos or via loop
 effects~\cite{Hall:1983id,Hempfling:1995wj,Nardi:1996iy}.
We will be constraining ourself to bilinear R-parity violation
(BRpV)~\cite{Kaplan:1999ds,Hirsch:2000ef,Diaz:2003as},
 thus breaking lepton number but not baryon
number. This model can be viewed as the effective theory of a
spontaneously broken $R$-parity
\cite{Santamaria:1987uq,Romao:1991ex,Nilles:1996ij,Kitano:1999qb}.
The BRpV neutrino
mass model has only a few free parameters and therefore is very
predictive.  Furthermore, in contrast to trilinear RPV neutrino mass
models, the constraints from LEP on the $R$-parity violating couplings
are automatically satisfied as the couplings are small due to the
requirement of explaining correctly neutrino data.

The RPV couplings giving rise to neutrino masses are also responsible
for the decay properties of the neutralino.  Therefore, an important
smoking gun signal for these models is the strong connection between
neutralino physics and neutrino mixing parameters.  Some ratios of the
branching ratios of the neutralino, or more generally the lightest
supersymmetric particle, are related to neutrino mixing
angles
\cite{Mukhopadhyaya:1998xj,Porod:2000hv,Hirsch:2002ys,Hirsch:2003fe}.
In particular, approximately the same
number of muons as taus are expected along with a $W$-boson because their
ratio is given by tangent of the nearly maximal atmospheric mixing
angle. By measuring the decay properties of the neutralino, which is
likely to be done by the LHC, a severe test of this model is possible.

Due to the smallness of the RPV couplings the neutralino will have a
long lifetime, but short enough for it to mainly decay within the
detectors at the LHC.
The prospects for collider discovery of RPV, responsible for the
neutrino masses and mixings, in mSUGRA have been thoroughly
studied~\cite{Barger:2001xe,Magro:2003zb,deCampos:2005ri,deCampos:2007bn,DeCampos:2010yu}.

Once this scenario is confirmed an important question will be how
well  the underlying parameters can be determined. There have
been several studies within the MSSM with conserved $R$-parity
\cite{Blair:2000gy,Blair:2002pg,Bechtle:2005vt,Lafaye:2007vs,%
Bechtle:2009ty,Adam:2010uz,Hirsch:2011cw}. In this paper we want to focus on
the determination of the RPV parameters taking into account
neutrino and LHC data. We will focus on a specific scenario
for the details but comment on how this can be extrapolated to
others. We will also work out which are the most sensitive
observables.

This paper is organized as follows: in the next section we will
recall briefly the main features of the BRpV model. In Section
\ref{sec:fit_procedure} we discuss the details of the fit procedure
and present the fit results in Section \ref{sec:results} and then
draw in Section \ref{sec:conclusions} our conclusions.

\section{Explicit bilinear \texorpdfstring{$R$}{R}-parity violation}
\label{sec:explicit_brpv}

The MSSM with explicit bilinear $R$-parity violation is specified by the
superpotential \cite{Diaz:1997xc}
\begin{gather}
  \Spot_{\mathrm{BRpV}} = \Spot_{\mathrm{MSSM}}
    + \epsilon_i \widehat{L}_i \widehat{H}_u \, ,
  \label{eq:brpv_superpotential}
\end{gather}
where the last term explicitly violates both $R$-parity and lepton number
in all three generations. In addition one has to add corresponding terms to
the soft SUSY breaking potential,
\begin{gather}
  V_{\mathrm{soft}}^{\mathrm{BRpV}} = V_{\mathrm{soft}}^{\mathrm{MSSM}} +
    B_i \epsilon_i \tilde{L}_i H_u \, ,
  \label{eq:brpv_soft_breaking_potential}
\end{gather}
which induce vacuum expectation values (VEVs) $v_i \Let
\Braket{\tilde{\nu}_i}$ for the sneutrinos. Using the tadpole
equations one can take the sneutrino VEVs as input instead of the
$B_i$ \cite{Hirsch:2004he}.

One important aspect of this model is that the \RpV{} terms give rise to
mixings between SM and SUSY particles. In the neutral
fermion sector the mixing between neutralinos and neutrinos leads to one
massive neutrino at tree-level while the other two neutrinos acquire masses
through loop corrections \cite{Hirsch:2000ef,Diaz:2003as,Hirsch:2004he}.
The effective neutrino mass matrix at tree- and one-loop level is given by
\begin{align}
  (m_{\mathrm{eff,LO}})_{ij}  &= a \Lambda_i \Lambda_j
  \label{eq:nu_mass_matrix_lo} \\
  (m_{\mathrm{eff,NLO}})_{ij} &= b \Lambda_i \Lambda_j
    + c (\Lambda_i \epsilon_j + \epsilon_i \Lambda_j)
    + d \epsilon_i \epsilon_j \, ,
  \label{eq:nu_mass_matrix_nlo}
\end{align}
where $a$, $b$, $c$, $d$ are functions of $R$-parity conserving parameters
and $\Lambda_i$ are the so called alignment parameters
\begin{equation}
\Lambda_i
= \mu v_i + v_d \epsilon_i\,.
\end{equation}
For example $a$ is given by
\begin{equation}
a = \frac{-(g^2 M_1 + g'{}^2 M_2)}{\sqrt{\Pi_{j=1}^4 m_{\tilde \chi^0_j}}}
\end{equation}
and $b$ is equal to $a$ plus radiative corrections.
 From the mixing matrix that diagonalizes
eq.~(\ref{eq:nu_mass_matrix_nlo}) one obtains expressions for the neutrino
mixing angles in terms of the \RpV{} parameters which
can approximately be expressed as  \cite{Hirsch:2000ef,Diaz:2003as}:
\begin{gather}
  \tan^2\theta_{23} \approx \left(\frac{\Lambda_2}{\Lambda_3} \right)^2
  \, , \qquad
  \tan^2\theta_{13} \approx \frac{\Lambda_1^2}{ \Lambda_2^2 + \Lambda_3^2}
  \, , \qquad
  \tan^2\theta_{12} \approx \left(\frac{\widetilde{\epsilon_1}}{
    \widetilde{\epsilon_2}}\right)^2
  \, ,
  \label{eq:nu_mixing_angles_approx}
\end{gather}
where $\widetilde{\epsilon_i} = V_{\mathrm{tree},ij} \epsilon_j$ and
$V_{\mathrm{tree}}$ diagonalizes the tree-level neutrino mass matrix
eq.~(\ref{eq:nu_mass_matrix_lo}). As one can
see from eqs.~(\ref{eq:nu_mass_matrix_nlo}) and
(\ref{eq:nu_mixing_angles_approx}) both the neutrino masses and the mixing
angles are predicted in terms of the \RpV{} parameters. It turns
out that the approximation for $\tan^2\theta_{23}$ is relative
insensitive when going from tree-level to one-loop level. The one for
$\tan^2\theta_{13}$ can be quite sensitive to loop corrections,
in particular if $\epsilon_2 \Lambda_2 \epsilon_3 \Lambda_3 >0$.
\cite{Hirsch:2000ef,Diaz:2003as}. This will also manifest
itself later in the fits discussed in Section \ref{sec:results}.
The most important loop contributions are due to sbottom-bottom and
stau-tau loops \cite{Hirsch:2000ef,Diaz:2003as}
which are  in both cases  proportional
to $\epsilon_j \epsilon_k / |\mu|^2$. However, there
are also regions in parameter space where the mixing between sleptons
and the charged Higgs boson \cite{Diaz:2003as} and/or the
sneutrino-neutrino loops give large contributions \cite{Dedes:2006ni}.

The violation of $R$-parity also implies that the
lightest supersymmetric particle (LSP), which we assume to be the lightest
neutralino here, is no longer stable and typically decays inside the
detector once the parameters are adjusted to satisfy the neutrino
constraints \cite{Porod:2000hv,Hirsch:2002ys,Hirsch:2003fe,Hirsch:2005ag}.
The parameters that
determine the decay properties of the LSP are the same parameters that lead
to neutrino masses and oscillation which implies that there are correlations
between the neutralino branching ratios and the neutrino mixing angles
 \cite{Mukhopadhyaya:1998xj,Porod:2000hv}, e.\,g.\
\begin{gather}
  \frac{\mathrm{BR}(\neutralino_1 \to W^\pm \mu^\mp)}
       {\mathrm{BR}(\neutralino_1 \to W^\pm \tau^\mp)}
    \approx \tan^2\theta_{23}
  \label{eq:chi10_br_ratio}
\end{gather}
This can be most easily seen by performing first an approximate diagonalization
of the chargino and neutralino mass matrices  as in \cite{Hirsch:1998kc}.
The parts of the corresponding  mixing matrices responsible for the
mixing between neutrinos and neutralinos as well as charged leptons
and charginos can be expressed in terms of
\begin{eqnarray}
\frac{\epsilon_i}{\mu} \,,\hspace{2mm}
\frac{\Lambda_i}{\sqrt{\Pi_{j=1}^4 m_{\tilde \chi^0_j}}} \,,\hspace{2mm}
\frac{\Lambda_i}{m_{\tilde \chi^+_1}m_{\tilde \chi^+_2}} \,.
\label{eq:RPratios}
\end{eqnarray}
and enter also the corresponding couplings in the decays used in
eq.~(\ref{eq:chi10_br_ratio}).
Moreover, the decay length of the LSP is proportional to the
$R$-parity breaking parameters
 \cite{Porod:2000hv,deCampos:2005ri,DeCampos:2010yu}. All these facts
 can be used to determine these parameters once information on
 the $R$-parity conserving ones is available.

\section{Fit Procedure}
\label{sec:fit_procedure}

 LHC will provide first information on the SUSY parameters once
the corresponding signals are observed. However, it will be
unlikely that the complete spectrum will be discovered and, thus,
the first parameter fits will be performed within specific high scale
models \cite{Bechtle:2005vt,Lafaye:2007vs,Bechtle:2009ty,Adam:2010uz}.
 Therefore,
the fits presented in this paper are performed in mSUGRA\footnote{Taking
mSUGRA is not crucial, as the explanation of neutrino data
does not depend  on this assumption and can equally well be
explained in GMSB \cite{Hirsch:2005ag}, AMSB
 \cite{Diaz:2002ij,deCampos:2004iy,deCampos:2008av} or the general
MSSM \cite{Porod:2000hv,Hirsch:2003fe}.} models which are
augmented by bilinear \RpV{} parameters at the electroweak scale.%
\footnote{This model is sometimes called RmSUGRA or BRpV-mSUGRA in the
literature.} These models therefore have eleven free parameters, namely the
five mSUGRA parameters $m_0$, $M_{1/2}$, $A_0$, $\tan\beta$, and
$\Sign{\mu}$ and the six bilinear \RpV{} parameters $\epsilon_i$, and
$\Lambda_i$ (or $v_i$ respectively). For the theoretical predictions of the
neutrino oscillation data, the LSP decay properties, and LHC/ILC
observables, \SPheno{}\footnote{The latest \SPheno{} version can
be obtained
from: \url{http://physik.uni-wuerzburg.de/~porod/SPheno.html}}%
\cite{Porod:2003um} version 3.0.beta54 has been used.

In order to measure the agreement between the data and the model for a
particular choice of parameters a simple $\chi^2$ function is used,
\begin{align}
  \chi^2(\bm{a}) = \sum_{i}
    \left(y_i - f_i(\bm{a}) \right)^2 / \sigma_i^2 \, ,
  \label{eq:chi_square}
\end{align}
where the $y_i$ are data points with their associated uncertainties or
experimental errors $\sigma_i$ and the $f_i(\bm{a})$ are theoretical
predictions for these data points at the point $\bm{a}$ in parameter space.
The data points used for the fits were also calculated by \SPheno{} for a
specific mSUGRA point, where the \RpV{} parameters are not explicitly
specified but are calculated iteratively such that the predicted neutrino
mixing angles and squared mass differences lie in the $3 \sigma$ confidence
region as given in~\cite[Table A1]{Schwetz:2008er}. If the \RpV{}
parameters as determined by \SPheno{} are denoted by $\bm{\widehat{a}}$,
then the data points are equal to their predictions at the point
$\bm{\widehat{a}}$ where the \RpV{} parameters are explicitly specified,
i.\,e.\ $y_i = f_i(\bm{\widehat{a}})$.

Since the data points used in the fits are themselves theoretical
predictions, the absolute minimum of eq.~(\ref{eq:chi_square}) is trivially
given by $\chisqmin \Let \chi^2(\bm{\widehat{a}}) = 0$. Instead of finding
the parameter point with the best goodness of fit, the purpose of data
fitting is then to estimate the parameter errors based on the uncertainties
$\sigma_i$, or to locate other minima in a multimodal $\chi^2$-landscape
that have equally good $\chi^2$-values. In order to find all minima in a
specific region of parameter space and to determine the boundaries of
$\chi^2(\bm{a})$ for the error estimation, the \Minuit{}
\Migrad{}\cite{James:1975dr} optimization algorithm was used repeatedly at
random starting points in parameter space. This procedure gives a good
coverage of the $\chi^2$-landscape and finds with high probability all
minima which lie in the region that is bounded by the starting points.
As interface between \SPheno{} and \Minuit{} the general purpose fitting
program \Kaimini{}\footnote{The latest \Kaimini{} version can be obtained
from: \url{https://github.com/fthomas/kaimini}}, which provides different
deterministic and stochastic optimization algorithms and works with any
program that implements the SUSY Les Houches Accord
(SLHA)\cite{Allanach:2008qq,Skands:2003cj}, is used.

\section{Results}
\label{sec:results}

\subsection{Fit Setup}
\label{ssec:fit_setup}

Various fits with different free parameters and data points were carried
out. The following subsections discusses fits where the \RpV{} parameters
are free parameters and different observables that depend on these
parameters (that are the neutrino oscillation data and the $\neutralino_1$
decay properties) are used as data points. For each of these combinations
three fits are performed which differ with respect to the mSUGRA
parameters being fixed or free parameters of the fit and the set of
corresponding data points. In the first setup the mSUGRA parameters are
fixed and only the neutrino and/or neutralino observables are used as data
points. In the second and third
setup the mSUGRA parameters $m_0$, $M_{1/2}$, $A_0$ and $\tan\beta$ are
free parameters in addition to the \RpV{} parameters. The additional data points of
the second setup are the ``edge variables'' $(m^2_{ll})^\mathrm{edge}$,
$(m^2_{qll})^\mathrm{edge}$, $(m^2_{qll})^\mathrm{thres}$,
$(m^2_{bll})^\mathrm{thres}$, $(m^2_{ql})_\mathrm{min}^\mathrm{edge}$,
and $(m^2_{ql})_\mathrm{max}^\mathrm{edge}$, where the predicted relative
uncertainties were taken from \cite[Table 5.13]{Weiglein:2004hn}. This
setup is denoted as ``LHC''. The third setup includes in addition to the
edge variables the masses of the $\neutralino_i$, $\chargino_i$, sleptons,
and the $\sstop_1$ assuming that are measured at a prospective future
$e^+ e^-$ linear collider such as ILC or CLIC.
The relative uncertainties of these
observables were taken from \cite[Table 5.14]{Weiglein:2004hn}. This setup
is denoted as ``LHC+ILC''. To be conservative the total uncertainties in
both setups were obtained by summing statistic and systematic
uncertainties linearly. We note, that taking the relative uncertainties
for the collider observables
equal for different points of parameter space is a strong assumption
which would have to be confirmed by individual studies which are however
not available in the literature. However, as we will see below the
uncertainties on the RPV parameters are dominated by the uncertainties
of the measurements of neutralino decay branching ratios and that
the uncertainties on the measurements of masses and edge variables are
only sub-dominant.

Fits for the various setups
 were performed for different SUSY benchmark points,
including SPS~1a$^\prime$~\cite{AguilarSaavedra:2005pw}, 1a, 1b, and 3
\cite{Allanach:2002nj}. In this paper we discuss in detail as a typical
example SPS~1a$^\prime$ working out the main features as most
experimental studies for the collider observables
used have been performed for points close
by and, thus, the assumption of using the same relative
uncertainties can be more easily justified. The results
for the other study points and additional plots for SPS~1a$^\prime$
are given on our web page \cite{webpage:2011}. We are aware that
SPS~1a$^\prime$ is potentially excluded by recent LHC data
\cite{Khachatryan:2011tk,Aad:2011hh} although the corresponding
searches have been performed for the $R$-parity conserving case.
However, we have checked that qualitative
features do not depend on the point under study, e.\,g.\ they are
the same for SPS~3 which is not excluded by existing data.

\subsection{Neutrino oscillation data}
\label{ssec:results_nu_id}

The free parameters of these fits are the six \RpV{} parameters while the
data points are the neutrino oscillation data, which are the two mass
squared differences $\Delta m^2_{\atm}$ and $\Delta m^2_{\sol}$ and the
three mixing angles $\tan^2 \theta_{\atm}$, $\tan^2 \theta_{\sol}$, and
$\sin^2 \theta_{13}$. As uncertainties
 for the data points the mean value of the
upper and lower $1 \sigma$ errors from Table A1 of \cite{Schwetz:2008er}
were used. To get meaningful results from the fits, the number of degrees
of freedom, which is the number of data points minus the number of
parameters, must be equal to or greater than zero. Therefore the
$\neutralino_1$ decay width was used as additional data point
and we assumed that it can be measured with an accuracy
 of \unit{15}{\%} \cite{ros:2010}. We perform this fit separately to
 get an understanding how the different sectors contribute to a global
fit.

\begin{figure}[t]
  \centering
  \includegraphics[scale=.536]{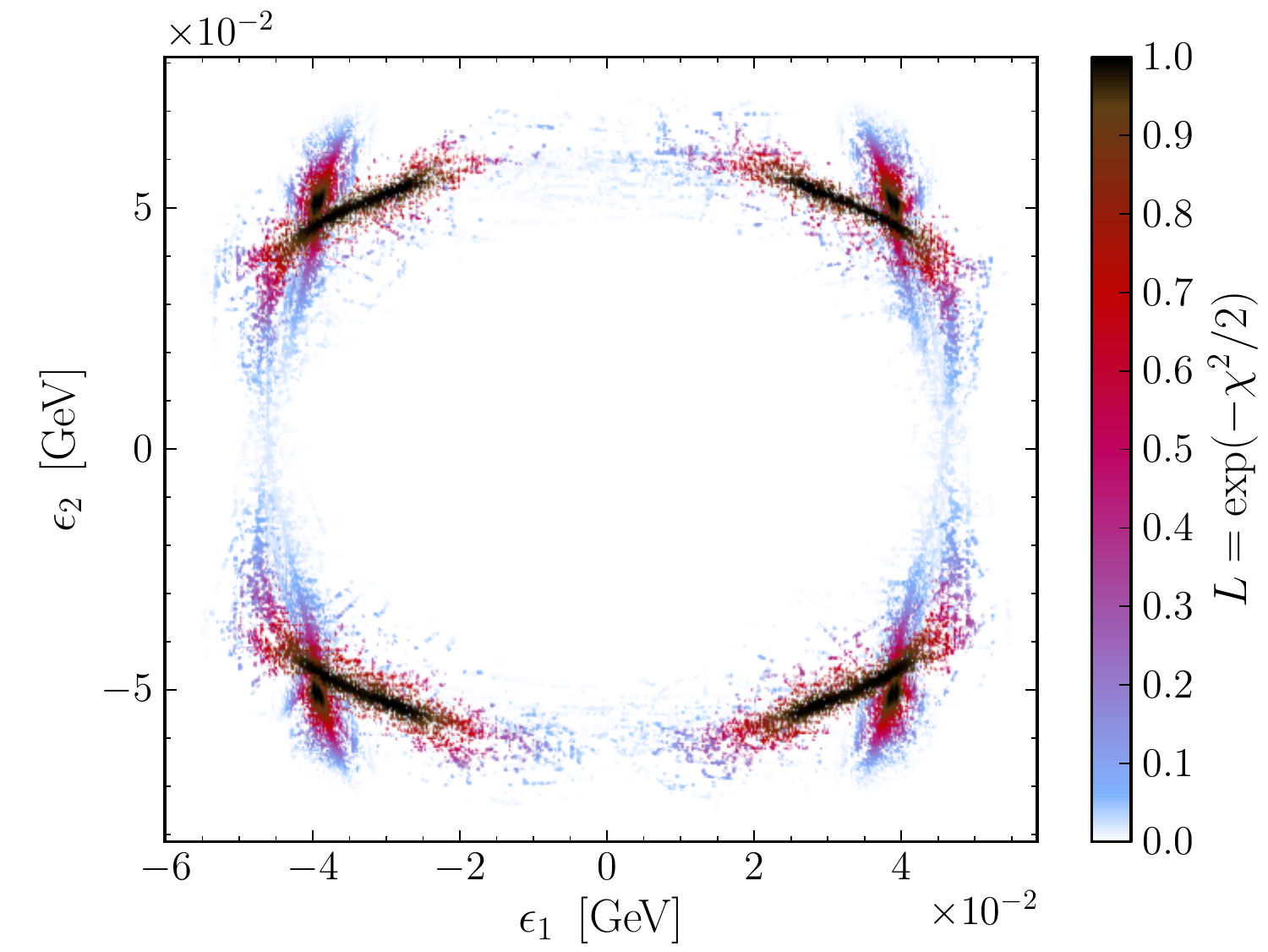}
  \includegraphics[scale=.536]{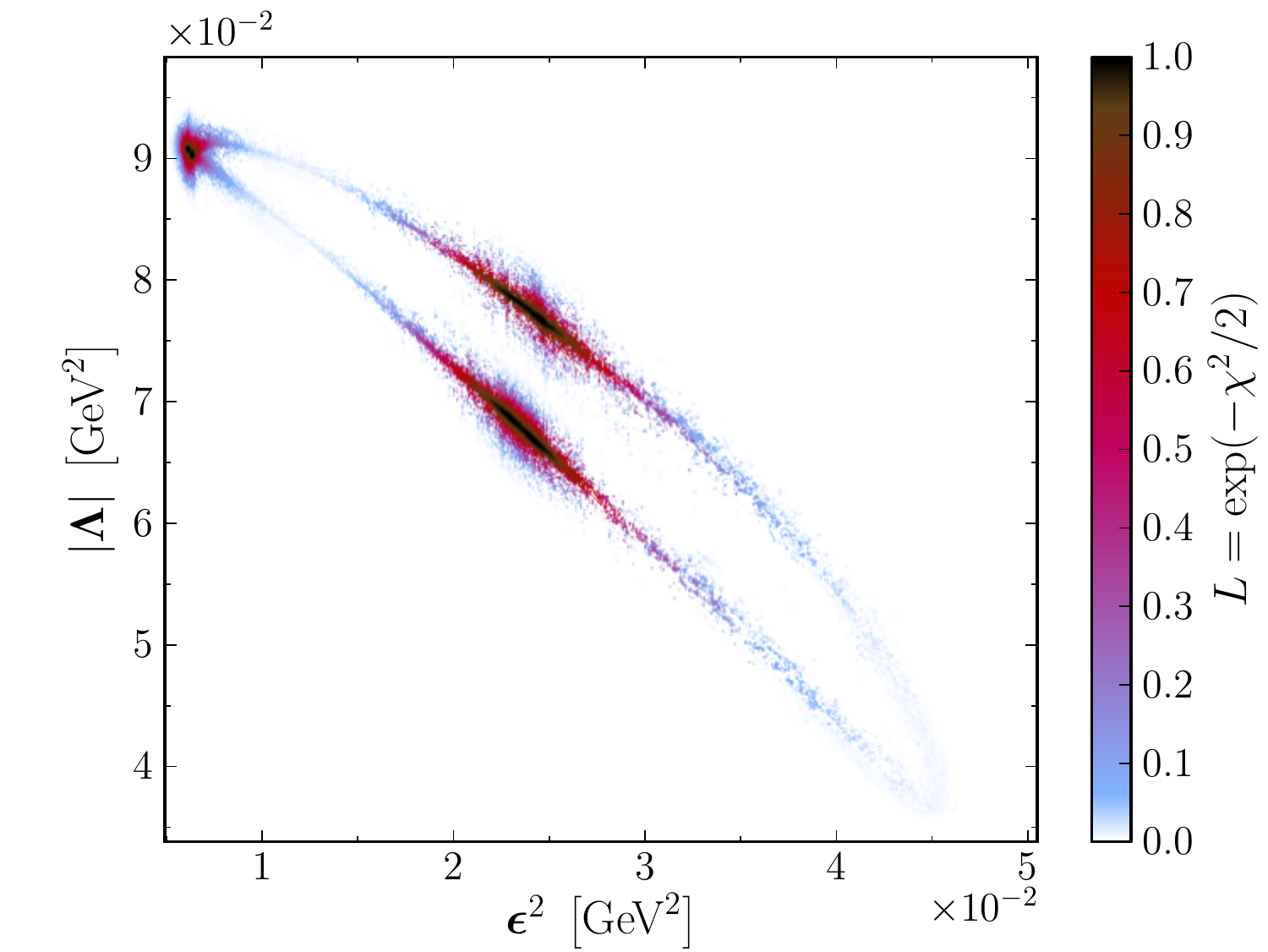} \\
  \includegraphics[scale=.536]{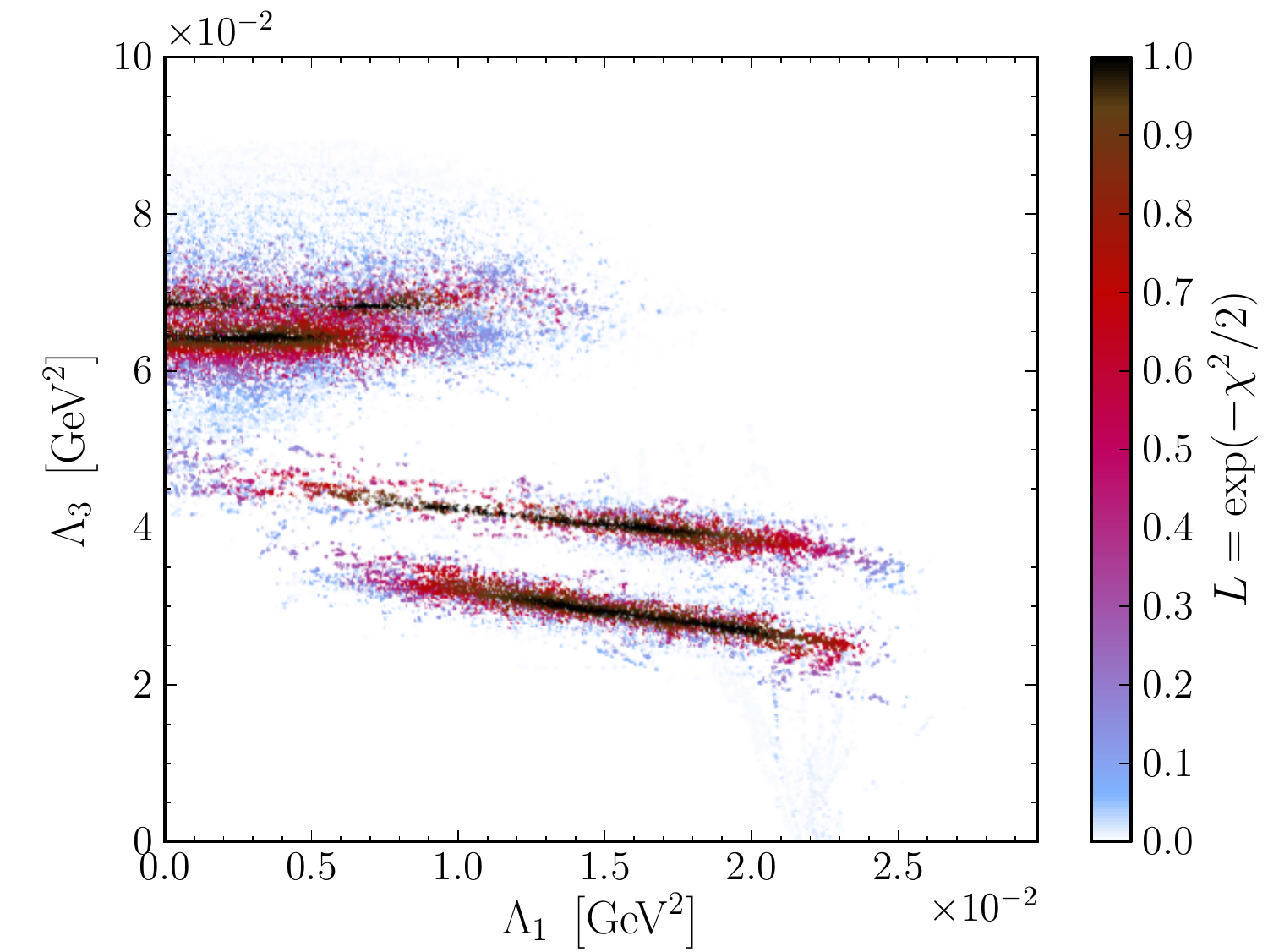}
  \includegraphics[scale=.536]{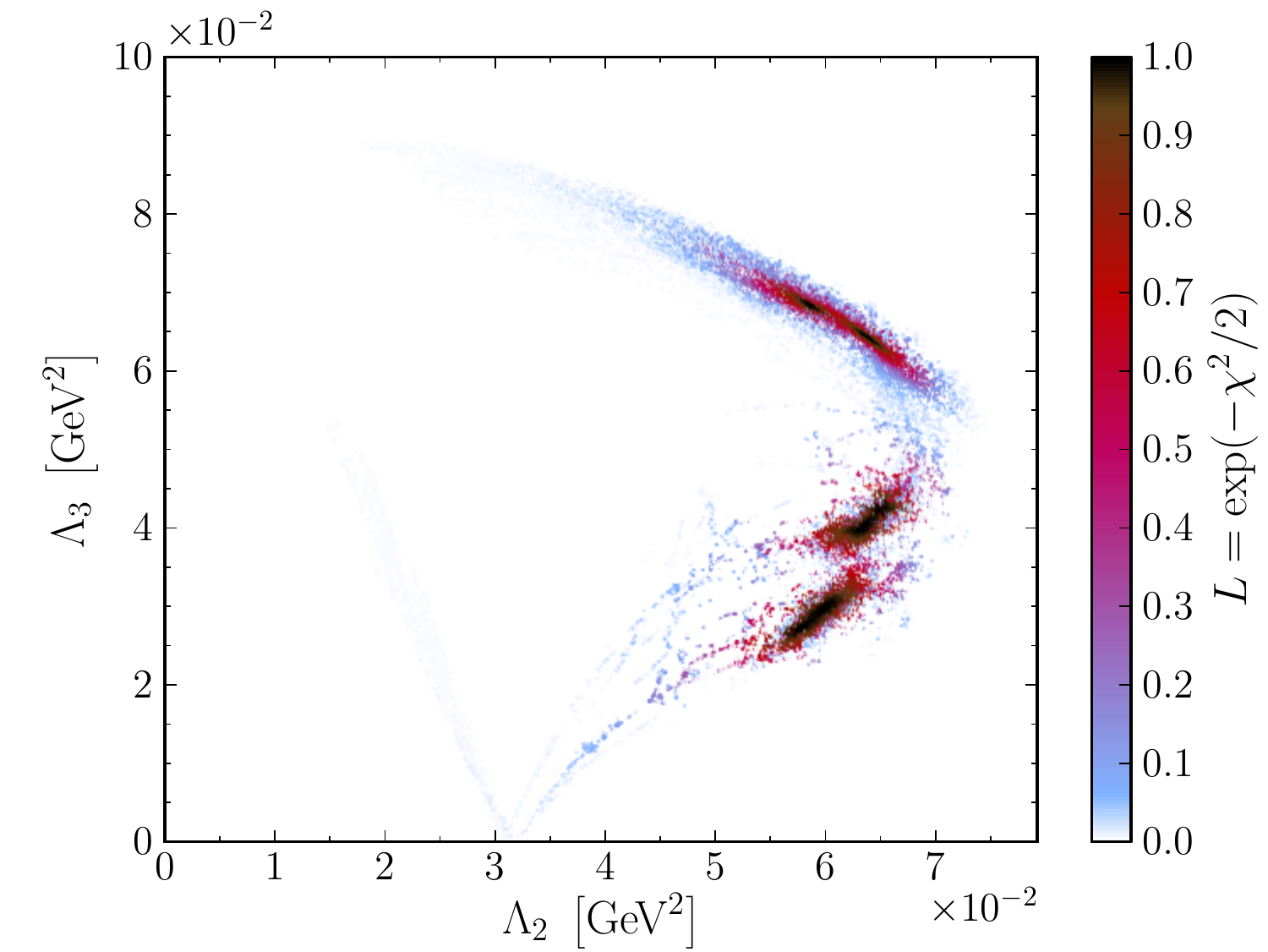}
\caption{Likelihood maps  of fits at
  SPS~1a$^\prime$ for various parameter combinations
using neutrino data and the neutralino decay width:
$\epsilon_1$-$\epsilon_2$ (upper left),
$\bm{\epsilon}^2$-$|\bm{\Lambda}|$ (upper right),
$\Lambda_1$-$\Lambda_3$ (lower left) and
  $\Lambda_2$-$\Lambda_3$ (lower right). These maps
  are projections of the six-dimensional parameter space onto the
  corresponding planes such that points with a higher likelihood cover
  points with a lower likelihood. The regions with maximum likelihood
  (which are colored black) correspond to minimal $\chi^2$-values.
  See section~\ref{ssec:results_nu_id} for further details.}
\label{fig:nu_id_l123}
\end{figure}

 One general result of these and all following fits
is that $\chi^2(\bm{a})$ is symmetric under the transformations $\Lambda_i
\rightarrow -\Lambda_i$ and
$\epsilon_i \to - \epsilon_i$,
i.\,e.\ for a given minimum there are several other
minima with the same goodness of fit which only differ in the signs of the
\RpV{} parameters. Independent of the specific mSUGRA point, the
$\chi^2$-landscape of these fits is rather complex with 8 minima per sign
combination of the $\Lambda_i$, resulting in 64 minima in total and each
of them having $\chi^2$-values of less than $10^{-5}$. For a specific
choice of signs for the $\Lambda_i$,
the 8 minima divide into two classes of 4 minima each
which differ by the importance of their individual loop contributions.
These minima can be differentiated by the sign of the ratio $\signcond$
\cite{Hirsch:2000ef,Diaz:2003as}.
For $\signcond < 0$, which is realized in the lower two plots
of Figure \ref{fig:nu_id_l123} for
$\Lambda_3 \lsim 0.05$~GeV$^2$,
 the most important contributions to the solar masses
and mixings come from bottom/sbottom and chargino/charged-scalar loops,
while for minima where $\signcond > 0$ (regions with
$\Lambda_3 \gsim 0.05$~GeV$^2$)
the corrections from bottom/sbottom
loops are dominant. In the latter case also
eqs.~(\ref{eq:nu_mixing_angles_approx}) get sizable corrections,
in particular $\tan^2 \theta_{13}$.
This can also be inferred from
the right upper plot of Figure~\ref{fig:nu_id_l123}. In
the region in the upper left corner the atmospheric neutrino mass
scale are essentially given by the tree level whereas in the lower
two regions sizeable loop corrections are needed.
 The ratio
  $\bm{\epsilon}^2 / |\bm{\Lambda}|$ gives the importance of the loop
  contributions relative to the tree-level-induced neutrino masses
  \cite{Hirsch:2000ef}.
In the left upper
plot we show preferred regions in the $\epsilon_2$-$\epsilon_3$ plane
to demonstrate the sign ambiguities of the underlying parameters.
Note, that in the  lower two  plots only
one quadrant is shown and the remaining ones can be obtained by
mirroring similar to the upper left.

The absolute uncertainties on these parameters depend clearly on
the $R$-parity conserving parameters but the relative uncertainties
are fairly insensitive to the $R$-parity conserving parameters. However,
up to now we have ignored their uncertainties as we only took neutrino
data into account. In the next sections we will add collider
observables to allow also for a variation of these parameters.

\subsection{Neutralino decay properties}
\label{ssec:results_chi10}

We now investigate how the neutralino decay properties
can be used to get information on the \RpV{} parameters
without using neutrino data.
We will use
 the $\neutralino_1$ decay width and branching ratios as
data points for the fits but keep the other collider
observables fixed.
Of the branching ratios only those were taken into account
whose values exceeded \unit{0.01}{\%}. The relative uncertainties of the
branching ratios were assumed to be $\Delta\mathrm{BR}_i/\mathrm{BR}_i =
2/\sqrt{N_i}$ with $N_i = 2 \cdot 10^6 \ \mathrm{BR}_i$ for a LHC
integrated luminosity of \unit{100}{fb^{-1}} \cite{AguilarSaavedra:2005pw}.
In the uncertainties of branching ratios whose final state contained quarks
or $\tau$ leptons we reduced the corresponding number $N_i$
to $N_i/10$ in order to take potential uncertainties
in the jet reconstruction into account. In principle this has to be checked
by detailed Monte Carlo studies which are however beyond the scope of
this article.

\begin{figure}[t]
  \centering
  \includegraphics[scale=.536]{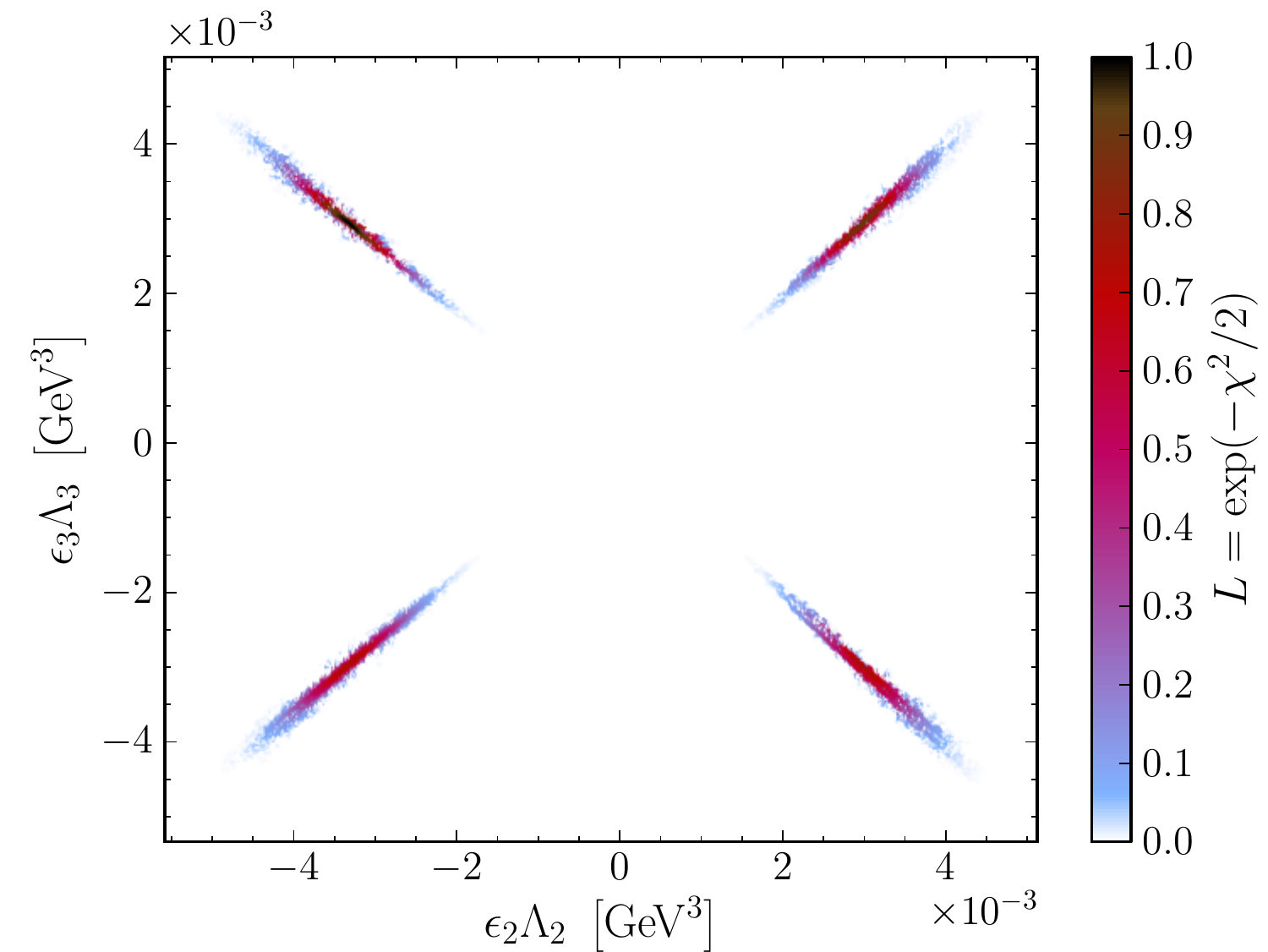}
\caption{Likelihood map of the $\epsilon_2\Lambda_2$-$\epsilon_3\Lambda_3$
  plain of fits at SPS~1a$^\prime$ with the $\neutralino_1$ decay
  properties (compare subsection~\ref{ssec:results_chi10}). The plot shows
  that the distinct regions which correspond to different sets of open
  decay channels have different signs for $\epsilon_2\Lambda_2$ and
  $\epsilon_3\Lambda_3$ and that for the global minimum
  $\epsilon_2\Lambda_2 < 0$ and $\epsilon_3\Lambda_3 > 0$. A fit with the
  neutrino and neutralino data combined will therefore favor points with
  this sign combinations.}
\label{fig:chi10_l2l3_signcond}
\end{figure}

\begin{figure}[t]
  \centering
  \includegraphics[scale=.536]{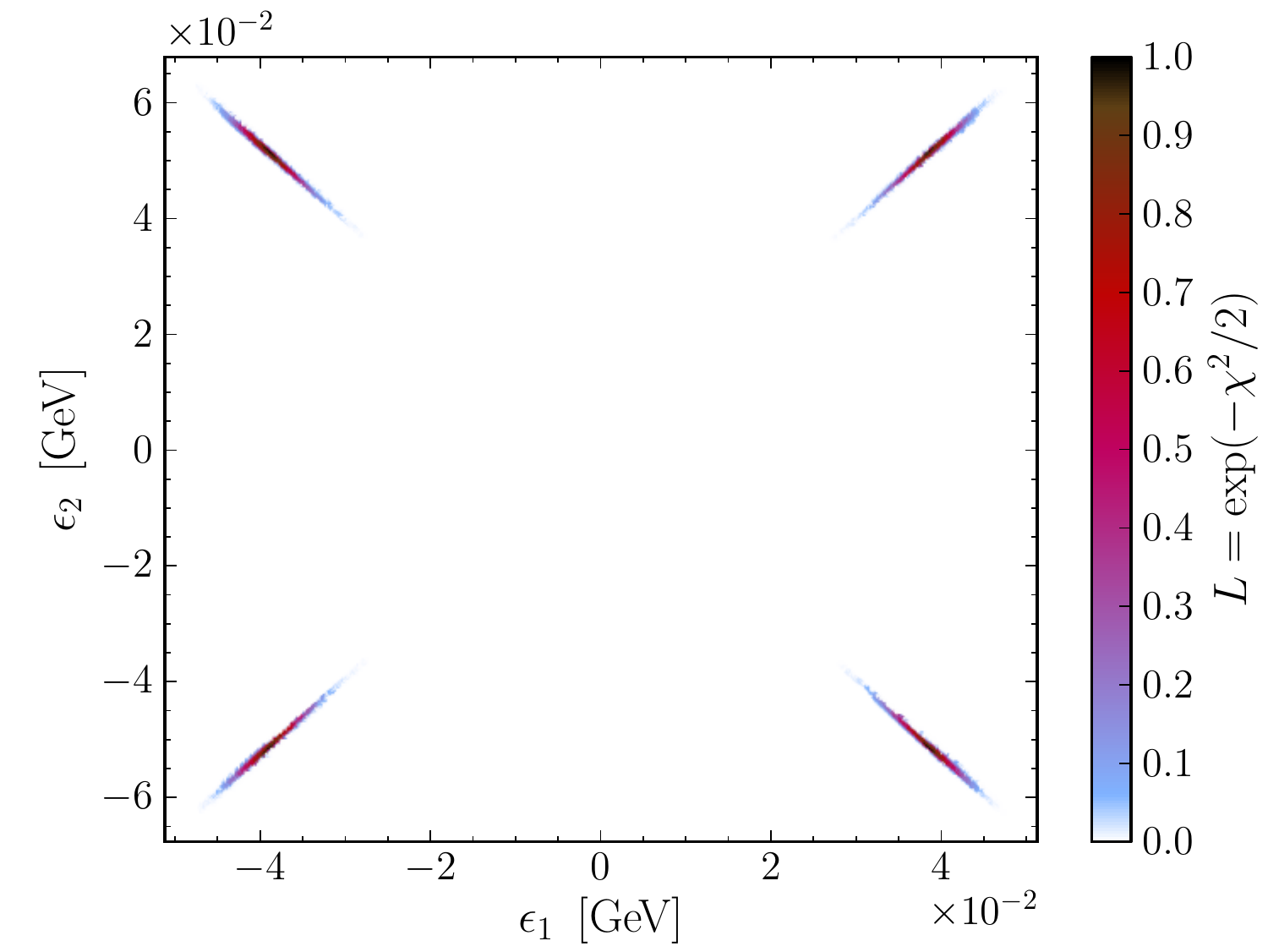}
  \includegraphics[scale=.536]{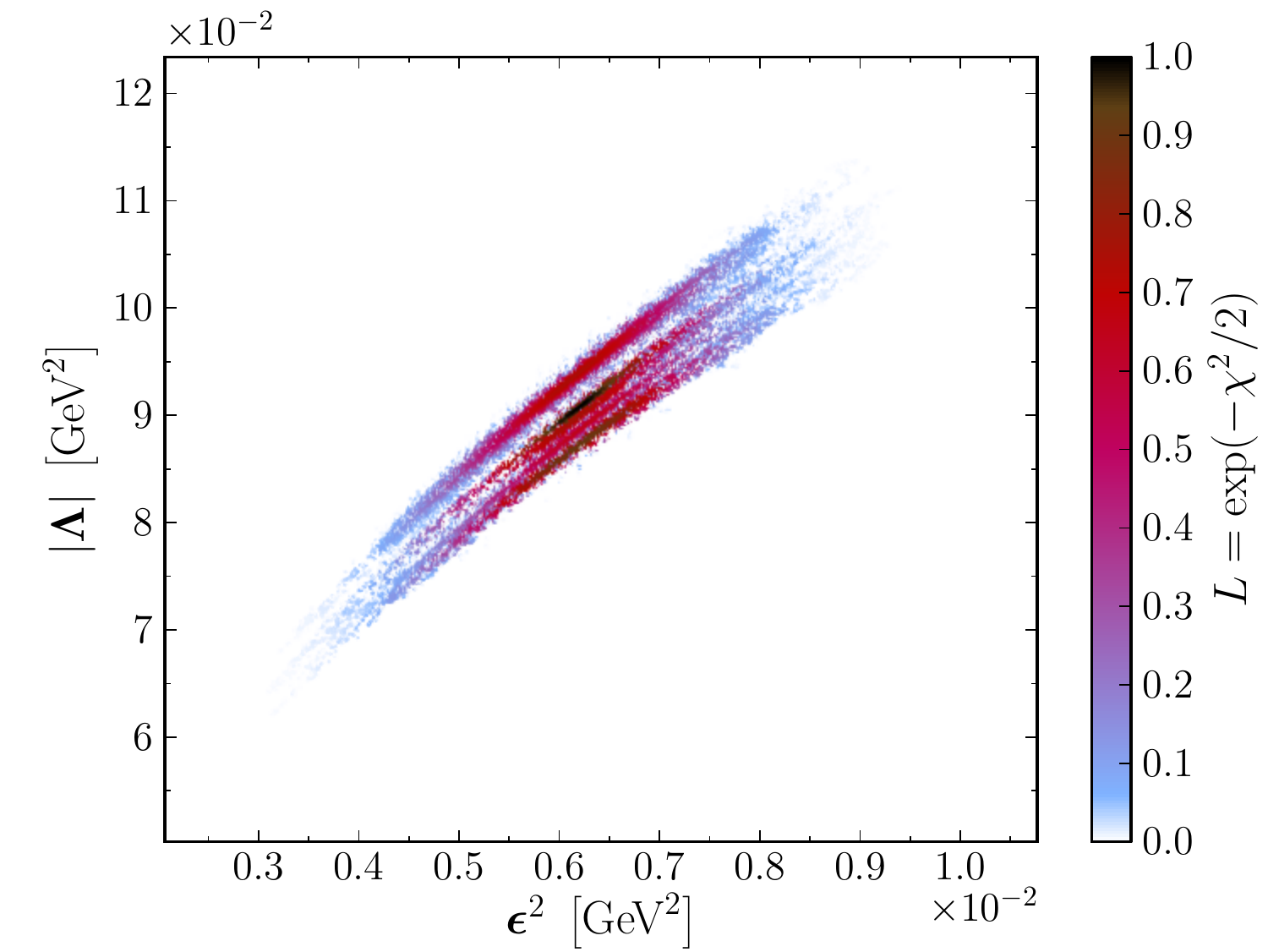} \\
  \includegraphics[scale=.536]{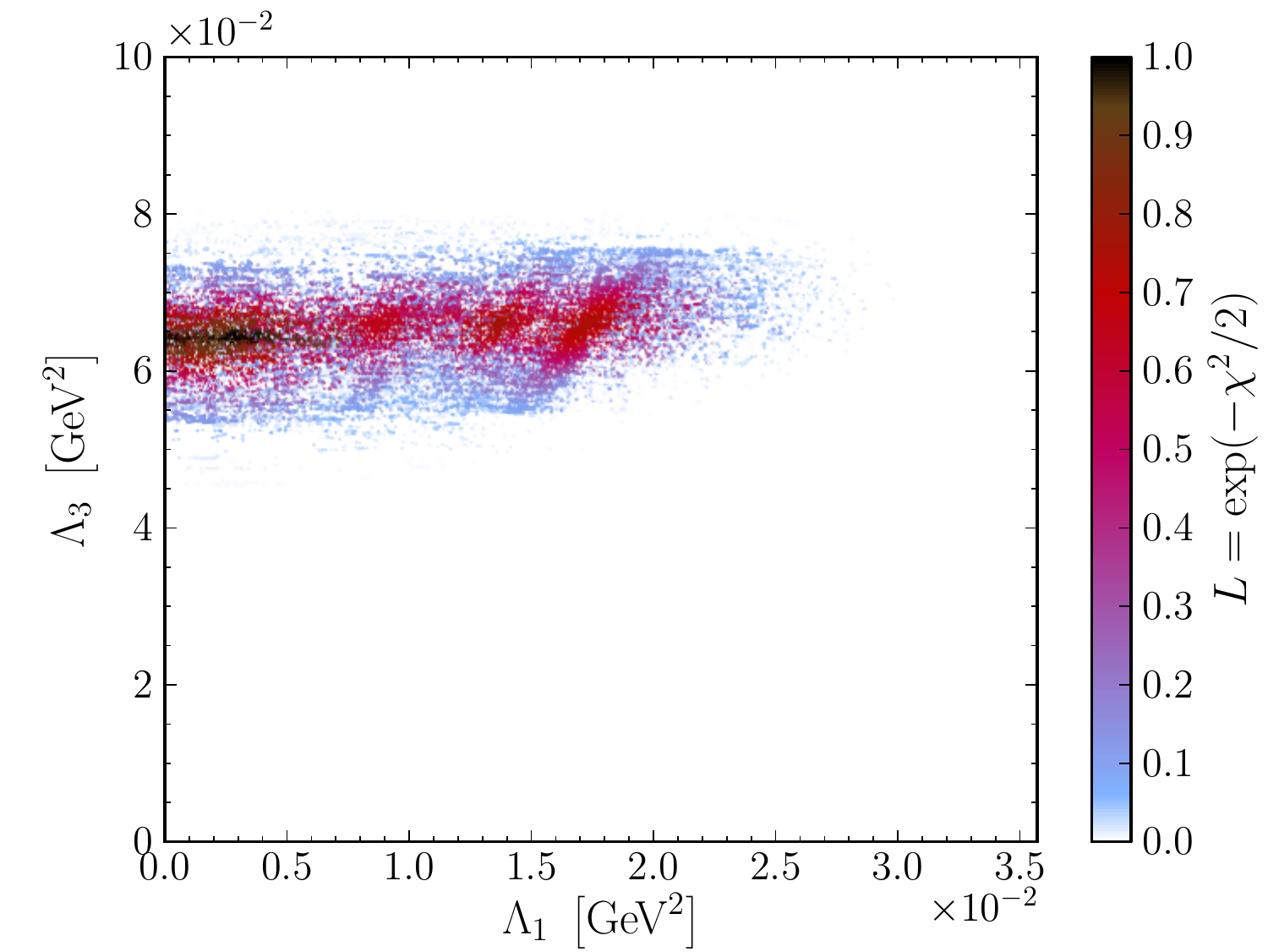}
  \includegraphics[scale=.536]{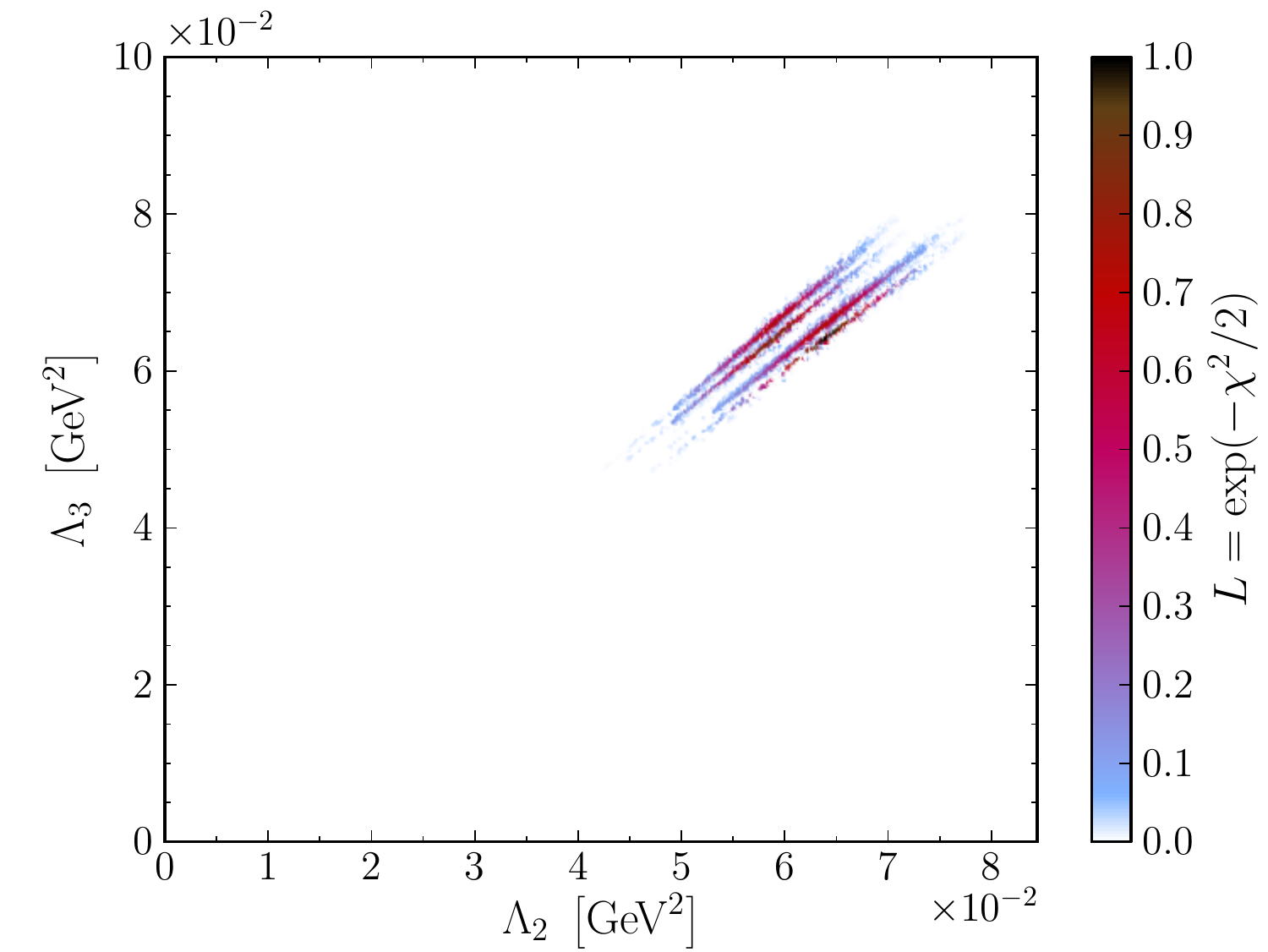}
\caption{Likelihood maps  of fits at
  SPS~1a$^\prime$ for various parameter combinations
using  neutralino decay width and neutralino branching ratios:
$\epsilon_1$-$\epsilon_2$ (upper left),
$\bm{\epsilon}^2$-$|\bm{\Lambda}|$ (upper right),
$\Lambda_1$-$\Lambda_3$ (lower left) and
  $\Lambda_2$-$\Lambda_3$ (lower right). In the lower two only
  one quadrant is shown, the others can be obtained by mirroring
  with respect to the axes.
  These maps
  are projections of the six-dimensional parameter space onto the
  corresponding planes such that points with a higher likelihood cover
  points with a lower likelihood. The regions with maximum likelihood
  (black) correspond to minimal $\chi^2$-values.
  See section~\ref{ssec:results_chi10} for further details.}
\label{fig:chi10_l123}
\end{figure}

As already mentioned in subsection~\ref{ssec:results_nu_id}, the
$\chi^2(\bm{a})$ of this fit is also symmetric under sign transformations
of the $\Lambda_i$, i.\,e.\
 also the decay properties of the $\neutralino_1$
do not depend on the sign of the alignment parameters. However, as can
be seen in Figure \ref{fig:chi10_l2l3_signcond} looking at the
parameter combinations $\epsilon_2 \Lambda_2$ and $\epsilon_3 \Lambda_3$
one can identify a preferred quadrant, in this case the upper left,
where all points with likelihood larger than 0.9 are located.

Compared to the
previous fits, the regions with a high goodness of fit is reduced which is
partly due to the higher number of degrees of freedom of this fit (13 data
points and 6 free parameters). There are now also distinct regions with a
high goodness of fit (each with its own local minimum) that correspond
to different sets of decay channels whose branching ratio exceeds the
aforementioned threshold of \unit{0.01}{\%}. These regions can be seen in
the two right plots of Figure~\ref{fig:chi10_l123}. Within this
setup the most important observables for the determination of the \RpV{}
parameters can be inferred by counting the frequency of those observables
with the highest $\chi^2$-contributions for different minima. For
SPS~1a$^\prime$ one finds for minima with $\chi^2 < 1$ that the four
branching ratios with the highest $\chi^2$-contributions are
\begin{gather}
  \BR{\neutralino_1}{\tau \tau \nu} \, , \quad
  \BR{\neutralino_1}{\tau \mu \nu} \, , \quad
  \BR{\neutralino_1}{\mu \mu \nu} \, , \quad
  \BR{\neutralino_1}{b \overline{b} \nu} \, .
\end{gather}
Note, that this finding depends to some extent on the parameter point
under study, e.\,g.\ for SPS~1b the most important branching ratios are
\begin{gather}
  \BR{\neutralino_1}{\tau \mu \nu} \, , \quad
  \BR{\neutralino_1}{\tau e \nu} \, , \quad
  \BR{\neutralino_1}{S^0_1 \nu} \, , \quad
  \BR{\neutralino_1}{b \overline{b} \nu} \, ,
\end{gather}
whereas for SPS~3 they are
\begin{gather}
  \BR{\neutralino_1}{\tau e \nu} \, , \quad
  \BR{\neutralino_1}{\mu \mu \nu} \, , \quad
  \BR{\neutralino_1}{\tau \mu \nu} \, , \quad
  \BR{\neutralino_1}{b \overline{b} \nu} \, .
\end{gather}
 Moreover, it is obvious by comparing Figures
 \ref{fig:nu_id_l123} and \ref{fig:chi10_l123} that neutrino data
and branching ratios give complementary information.

\subsection{Combination of neutrino and neutralino data}
\label{ssec:results_nu_id+chi10}

\begin{figure}[t]
  \centering
  \includegraphics[scale=.536]{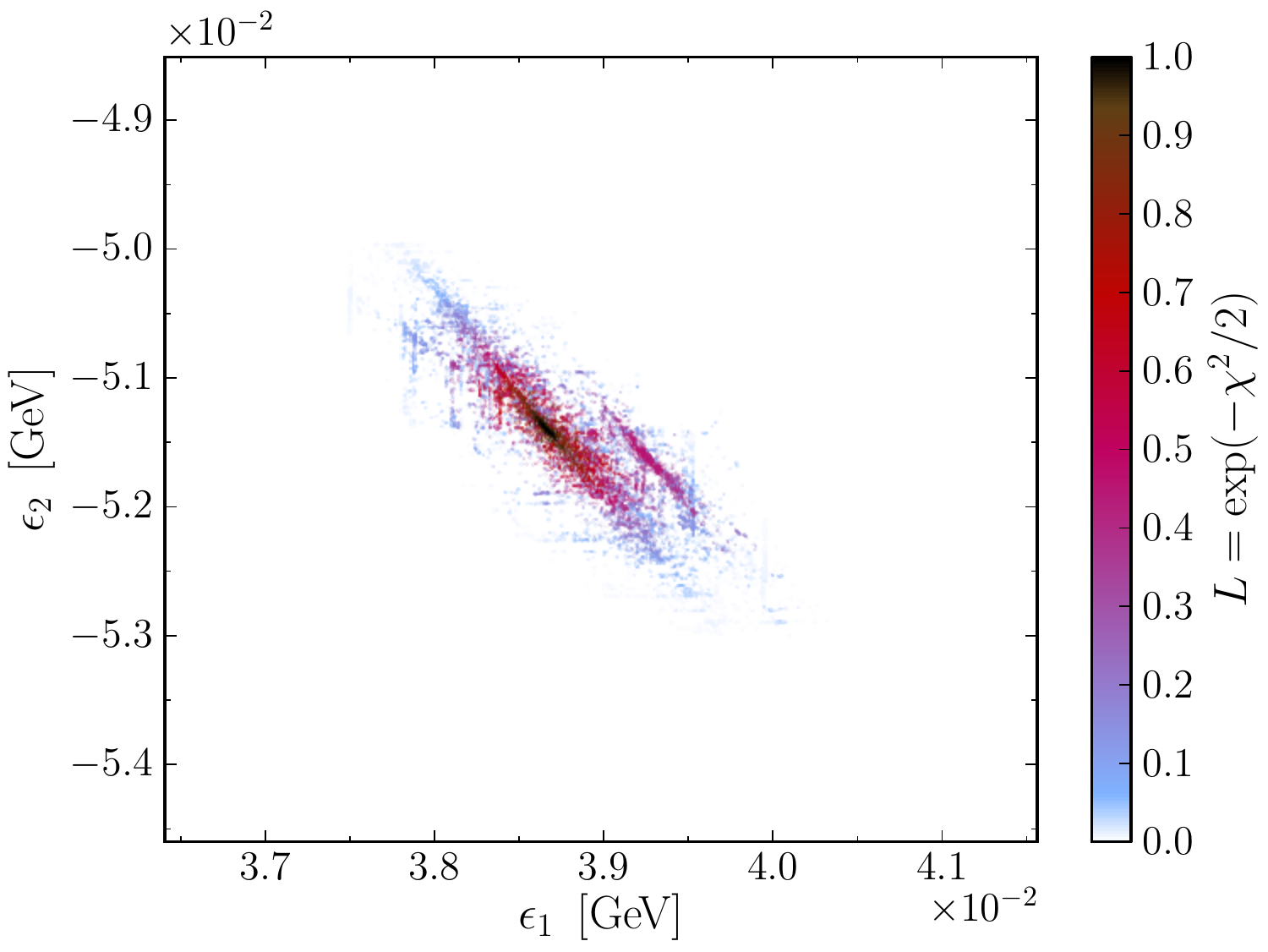}
  \includegraphics[scale=.536]{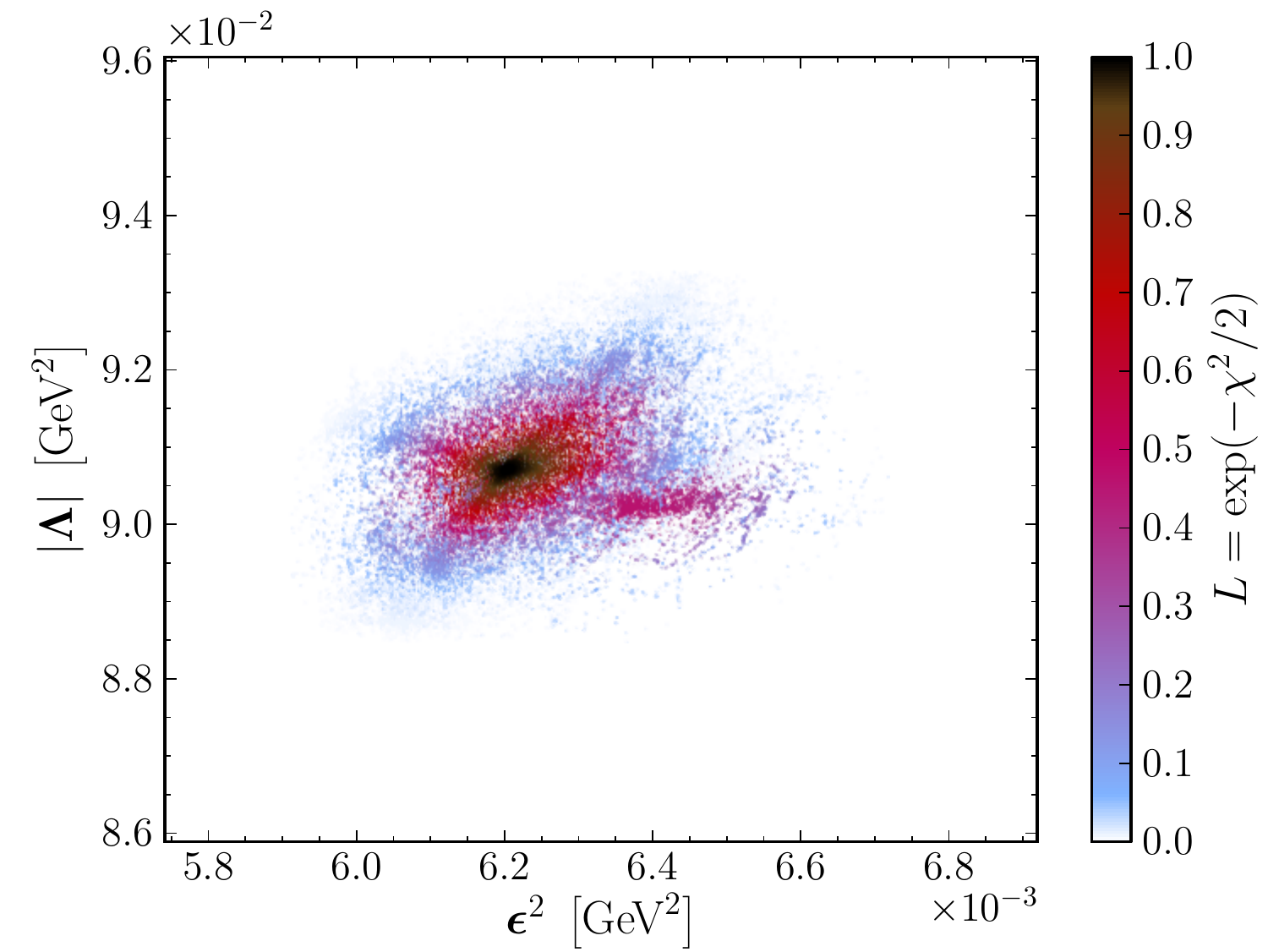} \\
  \includegraphics[scale=.536]{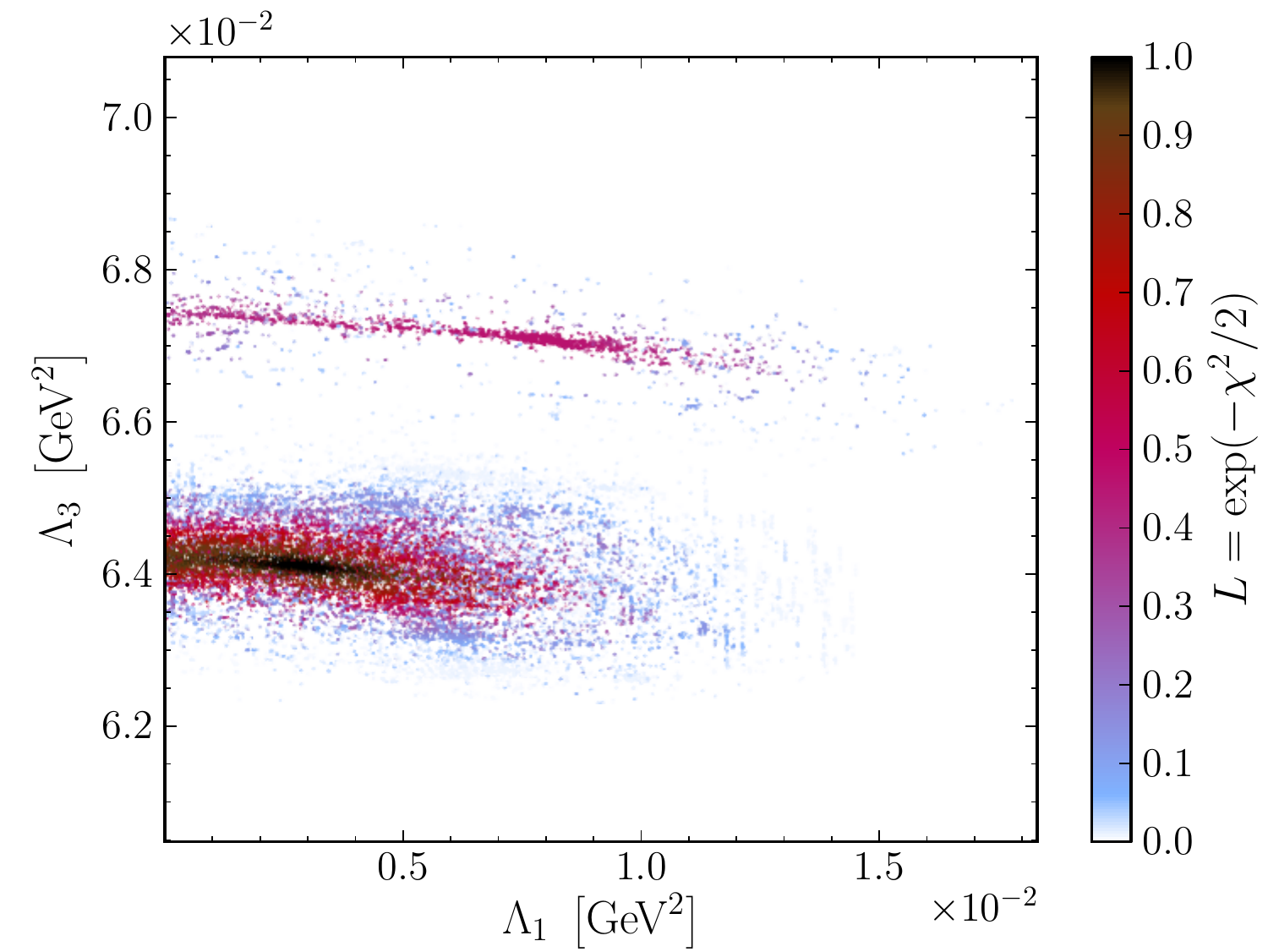}
  \includegraphics[scale=.536]{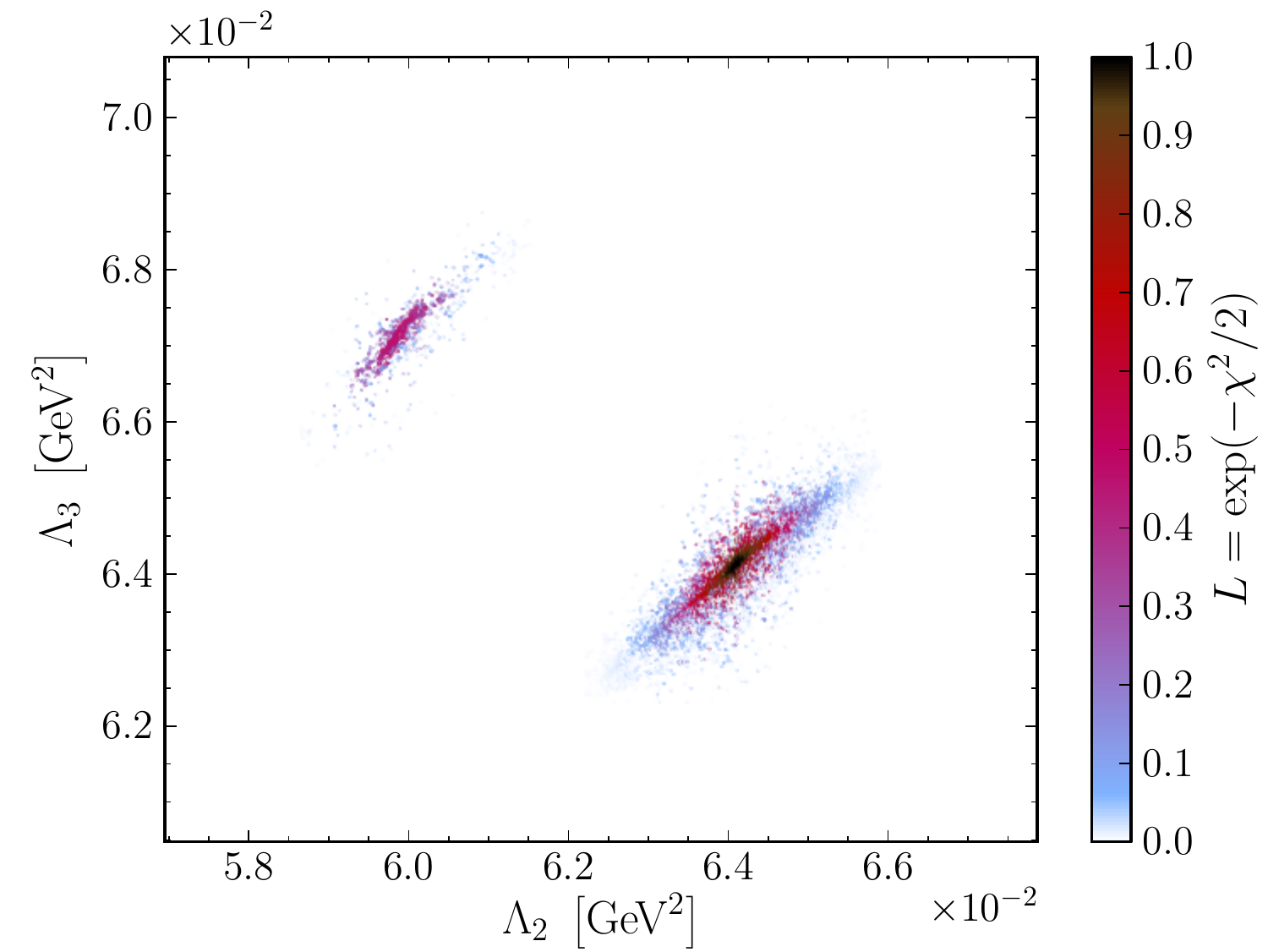}
\caption{Likelihood maps  of fits at
  SPS~1a$^\prime$ for various parameter combinations
using  neutralino decay width and neutralino branching ratios:
$\epsilon_1$-$\epsilon_2$ (upper left),
$\bm{\epsilon}^2$-$|\bm{\Lambda}|$ (upper right),
$\Lambda_1$-$\Lambda_3$ (lower left) and
  $\Lambda_2$-$\Lambda_3$ (lower right). In all but the upper right only
  one quadrant is shown, the others can be obtained by mirroring
  with respect to the axes.
  These maps
  are projections of the six-dimensional parameter space onto the
  corresponding planes such that points with a higher likelihood cover
  points with a lower likelihood. The regions with maximum likelihood
  (black) correspond to minimal $\chi^2$-values.
  See section~\ref{ssec:results_nu_id+chi10} for further details.}
\label{fig:nu_id+chi10_e1e2}
\end{figure}

In this section we are combining the previous two fits and
add also other observables: the \RpV{}
parameters are now fitted to the neutrino oscillation data and to the
$\neutralino_1$ decay properties. Due to the combination of the data
points, the regions with a high goodness of fit is significantly reduced
as can be seen in Figure \ref{fig:nu_id+chi10_e1e2}. Note,  the
different scaling
of the axes compared to the previous plots.
For SPS~1a$^\prime$ the parameter point $\widehat{\bm{a}}$ and all of its
reflections with respect to the $\Lambda_i$ axes are now the only minima
with $\chi^2 \simeq 0$. For these points $\epsilon_2\Lambda_2 < 0$ and
$\epsilon_3\Lambda_3 > 0$ hold. In the range $\chi^2 < 10$ there is only
one additional minimum with $\chi^2 \approx 1.5$ and for which
$\epsilon_2\Lambda_2 > 0$ and $\epsilon_3\Lambda_3 < 0$.

\begin{table}[t]
\begin{minipage}[b]{0.45\linewidth}\centering
\begin{tabular}{lccc}
  \addlinespace
  \multicolumn{4}{c}{SPS~1a$^\prime$}\\
  \toprule
  parameter & best fit & $2 \sigma$ & $3 \sigma$ \\
  \midrule
$\epsilon_1 \ \left[\unit{10^{-2}}{GeV}\right]$ & $3.87 \pm 0.03$ & $\pm 0.07$ & $^{+0.11}_{-0.10}$ \\
\addlinespace
$\epsilon_2 \ \left[\unit{10^{-2}}{GeV}\right]$ & $-5.14 \pm 0.04$ & $\pm 0.09$ & $\pm 0.13$ \\
\addlinespace
$\epsilon_3 \ \left[\unit{10^{-2}}{GeV}\right]$ & $4.55 ^{+0.10}_{-0.08}$ & $^{+0.22}_{-0.19}$ & $^{+0.32}_{-0.29}$ \\
\addlinespace
$\Lambda_1 \ \left[\unit{10^{-1}}{GeV}^2\right]$ & $0.03 ^{+0.04}_{-0.10}$ & $^{+0.08}_{-0.13}$ & $^{+0.11}_{-0.16}$ \\
\addlinespace
$\Lambda_2 \ \left[\unit{10^{-2}}{GeV}^2\right]$ & $6.41 ^{+0.05}_{-0.04}$ & $^{+0.11}_{-0.10}$ & $^{+0.16}_{-0.15}$ \\
\addlinespace
$\Lambda_3 \ \left[\unit{10^{-2}}{GeV}^2\right]$ & $6.41 \pm 0.05$ & $^{+0.12}_{-0.13}$ & $\pm 0.18$ \\
  \bottomrule
\end{tabular}
\end{minipage}
\begin{minipage}[b]{0.45\linewidth}\centering
\begin{tabular}{lccc}
  \addlinespace
  \multicolumn{4}{c}{SPS~3}\\
  \toprule
  parameter & best fit & $2 \sigma$ & $3 \sigma$ \\
  \midrule
$\epsilon_1 \ \left[\unit{10^{-2}}{GeV}\right]$ & $6.88 \pm 0.04$ & $^{+0.15}_{-0.14}$ & $\pm 0.18$ \\
\addlinespace
$\epsilon_2 \ \left[\unit{10^{-2}}{GeV}\right]$ & $-9.14 \pm 0.05$ & $\pm 0.17$ & $\pm 0.22$ \\
\addlinespace
$\epsilon_3 \ \left[\unit{10^{-2}}{GeV}\right]$ & $8.09 \pm 0.10$ & $\pm 0.33$ & $^{+0.42}_{-0.43}$ \\
\addlinespace
$\Lambda_1 \ \left[\unit{10^{-1}}{GeV}^2\right]$ & $0.05 ^{+0.02}_{-0.11}$ & $^{+0.05}_{-0.15}$ & $^{+0.06}_{-0.17}$ \\
\addlinespace
$\Lambda_2 \ \left[\unit{10^{-2}}{GeV}^2\right]$ & $10.84 \pm 0.06$ & $\pm 0.19$ & $\pm 0.24$ \\
\addlinespace
$\Lambda_3 \ \left[\unit{10^{-2}}{GeV}^2\right]$ & $10.84 \pm 0.06$ & $\pm 0.21$ & $^{+0.26}_{-0.27}$ \\
\bottomrule
\end{tabular}
\end{minipage}
\caption{Parameter point $\widehat{\bm{a}}$ at SPS~1a$^\prime$ and SPS~3 with
  $1\sigma$, $2\sigma$, and $3\sigma$ uncertainties (12 d.\,f.) from fits
  with the neutrino oscillation data and the $\neutralino_1$ decay
  properties.}
\label{tab:nu_id+chi10}
\end{table}

\begin{table}[t]
\begin{minipage}[b]{0.45\linewidth}\centering
\begin{tabular}{lccc}
  \addlinespace
  \multicolumn{4}{c}{SPS~1a$^\prime$}\\
  \toprule
  parameter & best fit & $2 \sigma$ & $3 \sigma$ \\
  \midrule
$\epsilon_1 \ \left[\unit{10^{-2}}{GeV}\right]$ & $3.87 \pm 0.03$ & $^{+0.08}_{-0.07}$ & $^{+0.12}_{-0.11}$ \\
\addlinespace
$\epsilon_2 \ \left[\unit{10^{-2}}{GeV}\right]$ & $-5.14 \pm 0.04$ & $\pm 0.09$ & $\pm 0.15$ \\
\addlinespace
$\epsilon_3 \ \left[\unit{10^{-2}}{GeV}\right]$ & $4.55 ^{+0.10}_{-0.08}$ & $^{+0.22}_{-0.19}$ & $^{+0.33}_{-0.30}$ \\
\addlinespace
$\Lambda_1 \ \left[\unit{10^{-1}}{GeV}^2\right]$ & $0.03 ^{+0.04}_{-0.09}$ & $^{+0.07}_{-0.14}$ & $^{+0.10}_{-0.16}$ \\
\addlinespace
$\Lambda_2 \ \left[\unit{10^{-2}}{GeV}^2\right]$ & $6.41 \pm 0.04$ & $\pm 0.10$ & $\pm 0.16$ \\
\addlinespace
$\Lambda_3 \ \left[\unit{10^{-2}}{GeV}^2\right]$ & $6.41 \pm 0.05$ & $\pm 0.12$ & $^{+0.17}_{-0.18}$ \\
  \bottomrule
\end{tabular}
\end{minipage}
\begin{minipage}[b]{0.45\linewidth}\centering
\begin{tabular}{lccc}
  \addlinespace
  \multicolumn{4}{c}{SPS~3}\\
  \toprule
  parameter & best fit & $2 \sigma$ & $3 \sigma$ \\
  \midrule
$\epsilon_1 \ \left[\unit{10^{-2}}{GeV}\right]$ & $6.88 \pm 0.05$ & $\pm 0.12$ & $\pm 0.20$ \\
\addlinespace
$\epsilon_2 \ \left[\unit{10^{-2}}{GeV}\right]$ & $-9.14 \pm 0.06$ & $\pm 0.15$ & $\pm 0.25$ \\
\addlinespace
$\epsilon_3 \ \left[\unit{10^{-2}}{GeV}\right]$ & $8.09 \pm 0.11$ & $\pm 0.26$ & $^{+0.42}_{-0.43}$ \\
\addlinespace
$\Lambda_1 \ \left[\unit{10^{-1}}{GeV}^2\right]$ & $0.05 ^{+0.02}_{-0.12}$ & $^{+0.06}_{-0.15}$ & $^{+0.09}_{-0.20}$ \\
\addlinespace
$\Lambda_2 \ \left[\unit{10^{-2}}{GeV}^2\right]$ & $10.84 \pm 0.06$ & $\pm 0.15$ & $^{+0.24}_{-0.25}$ \\
\addlinespace
$\Lambda_3 \ \left[\unit{10^{-2}}{GeV}^2\right]$ & $10.84 ^{+0.07}_{-0.06}$ & $^{+0.15}_{-0.17}$ & $^{+0.25}_{-0.26}$ \\
  \bottomrule
\end{tabular}
\end{minipage}
\caption{Parameter point $\widehat{\bm{a}}$ at SPS~1a$^\prime$ and SPS~3 with
  $1\sigma$, $2\sigma$, and $3\sigma$ uncertainties (30 d.\,f.)
  from fits with the ``LHC'' setup. }
\label{tab:lhc_nu_id+chi10}
\end{table}

\begin{table}[t]
\begin{minipage}[b]{0.45\linewidth}\centering
\begin{tabular}{lccc}
  \addlinespace
  \multicolumn{4}{c}{SPS~1a$^\prime$}\\
  \toprule
  parameter & best fit & $2 \sigma$ & $3 \sigma$ \\
  \midrule
$\epsilon_1 \ \left[\unit{10^{-2}}{GeV}\right]$ & $3.87 \pm 0.03$ & $^{+0.08}_{-0.07}$ & $\pm 0.12$ \\
\addlinespace
$\epsilon_2 \ \left[\unit{10^{-2}}{GeV}\right]$ & $-5.14 \pm 0.04$ & $\pm 0.09$ & $\pm 0.15$ \\
\addlinespace
$\epsilon_3 \ \left[\unit{10^{-2}}{GeV}\right]$ & $4.55 ^{+0.10}_{-0.08}$ & $^{+0.22}_{-0.19}$ & $^{+0.34}_{-0.31}$ \\
\addlinespace
$\Lambda_1 \ \left[\unit{10^{-1}}{GeV}^2\right]$ & $0.03 ^{+0.04}_{-0.09}$ & $^{+0.07}_{-0.13}$ & $^{+0.10}_{-0.16}$ \\
\addlinespace
$\Lambda_2 \ \left[\unit{10^{-2}}{GeV}^2\right]$ & $6.41 ^{+0.04}_{-0.05}$ & $\pm 0.10$ & $^{+0.17}_{-0.16}$ \\
\addlinespace
$\Lambda_3 \ \left[\unit{10^{-2}}{GeV}^2\right]$ & $6.41 \pm 0.05$ & $^{+0.10}_{-0.11}$ & $^{+0.18}_{-0.17}$ \\
  \bottomrule
\end{tabular}
\end{minipage}
\begin{minipage}[b]{0.45\linewidth}\centering
\begin{tabular}{lccc}
  \addlinespace
  \multicolumn{4}{c}{SPS~3}\\
  \toprule
  parameter & best fit & $2 \sigma$ & $3 \sigma$ \\
  \midrule
$\epsilon_1 \ \left[\unit{10^{-2}}{GeV}\right]$ & $6.88 \pm 0.05$ & $\pm 0.12$ & $\pm 0.20$ \\
\addlinespace
$\epsilon_2 \ \left[\unit{10^{-2}}{GeV}\right]$ & $-9.14 \pm 0.06$ & $\pm 0.15$ & $^{+0.25}_{-0.24}$ \\
\addlinespace
$\epsilon_3 \ \left[\unit{10^{-2}}{GeV}\right]$ & $8.09 \pm 0.11$ & $\pm 0.26$ & $^{+0.42}_{-0.43}$ \\
\addlinespace
$\Lambda_1 \ \left[\unit{10^{-1}}{GeV}^2\right]$ & $0.05 ^{+0.02}_{-0.12}$ & $^{+0.05}_{-0.15}$ & $^{+0.09}_{-0.19}$ \\
\addlinespace
$\Lambda_2 \ \left[\unit{10^{-2}}{GeV}^2\right]$ & $10.84 \pm 0.06$ & $\pm 0.15$ & $^{+0.25}_{-0.24}$ \\
\addlinespace
$\Lambda_3 \ \left[\unit{10^{-2}}{GeV}^2\right]$ & $10.84 ^{+0.06}_{-0.07}$ & $^{+0.15}_{-0.16}$ & $^{+0.24}_{-0.26}$ \\
  \bottomrule
\end{tabular}
\end{minipage}
\caption{Parameter point $\widehat{\bm{a}}$ at SPS~1a$^\prime$ and SPS~3 with
  $1\sigma$, $2\sigma$, and $3\sigma$ uncertainties (46 d.\,f.)
  from fits with the ``LHC+ILC'' setup.}
\label{tab:ilc_nu_id+chi10}
\end{table}

Tables~\ref{tab:nu_id+chi10}-\ref{tab:ilc_nu_id+chi10} give the
uncertainties of the \RpV{} parameters inferred from these fits for the
``neutrino+neutralino data'', the ``LHC'', and the ``LHC+ILC'' setup for
SPS~1a$^\prime$ and SPS~3. The main uncertainties are due to the uncertainties
of the neutrino data
and the neutralino decay branching ratios a
s can bee seen by comparing Table~\ref{tab:nu_id+chi10}
with Tables~\ref{tab:lhc_nu_id+chi10} and \ref{tab:ilc_nu_id+chi10}.
 Except from $\Lambda_1$ the
  percentage errors of the \RpV{} parameters at the $3\sigma$-level is
  less than \unit{8}{\%}. $\Lambda_1$ is particularly difficult as
  the corresponding branching ratio $\tilde \chi^0_1 \to W^\pm e^\mp$
  is very small leading to the large uncertainties obtained.
Note, that within the top-down approach, this means fitting the
high scale mSUGRA parameters, there is hardly an effect on the
uncertainties of the RPV parameters when going from the LHC to ILC.
However, we do not expect significant differences when a fit of the
$R$-parity conserving parameters is performed at the electroweak
scale provided the neutralino and chargino masses can be measured
within \unit{1-2}{\%} which is feasible at the ILC
 \cite{Accomando:1997wt,Weiglein:2004hn}
or at CLIC \cite{Alster:2011he}.
For the masses of the staus, sbottoms and the corresponding mixing angle
one would need an accuracy of \unit{10-20}{\%} percent to keep the corresponding
uncertainties in the calculation of the neutrino masses and mixing
angles small enough so that they are sub-dominant compared to the
uncertainties due to the branching ratios uncertainties. An important
question will of course be on how much better the neutralino branching
ratios can be measured at the ILC or CLIC compared to LHC.

\section{Discussion and conclusions}
\label{sec:conclusions}

In this paper we have investigated the question how well
one can measure $R$-parity violating couplings using
neutrino data and future collider data. For simplicity
we have taken bilinear $R$-parity violating parameters
and fixed the $R$-parity conserving ones by imposing mSUGRA boundary
conditions. The latter choice is not important as the requirement
of explaining correctly neutrino data only fixes ratios of
$R$-parity violating parameters over $R$-parity conserving parameters.

For the fits we have taken the current experimental accuracies
on neutrino data and the expected accuracies on the measurements
of edge variables at the LHC and mass measurements at the ILC
as given by the corresponding collider studies. In addition we have
assumed that at the LHC the decay length of the lightest neutralino
can be measured within 15 percent and that the branching ratios
can be determined within twice the corresponding statistical
uncertainties. Under this assumptions we find that the expected
accuracies on the RPV parameters is of order one percent. However,
all fits show that there will be a sign ambiguity as all observables
considered are nearly the same under a sign change of the RPV parameters.
This ambiguities might be resolved by detailed studies of the lepton and
jet spectra of the individual decay channels of the lightest supersymmetric
particle, in the case under study the lightest neutralino.

The main source of the uncertainties are due to the
uncertainties on the neutralino branching ratios. This implies
that in a top-down approach, e.\,g.\ fitting the $R$-parity conserving
within a given high scale model such as mSUGRA, the  main information
is already obtained at the LHC even if the ILC measures the
SUSY spectrum more precisely.
However, improvements are expected if at
at ILC the neutralino branching ratios can be measured significantly
better than at the LHC as one would naively presume.
In case one fits the $R$-parity conserving parameters at the electroweak
scale, then our findings hold if the masses
 of neutralinos, charginos, staus and sbottoms as well as
 the corresponding mixing
 matrices can be determined precisely. To keep things at the level shown
 (up to about a factor 1.5 -- 2) one needs in case of the neutralino
 and charginos sectors
 precision in the percent-range, in stau and sbottom sector \unit{10}{\%} accuracies
implying the need of an ILC and most likely also a multi-TeV
$e^+ e^-$ collider such as  CLIC.

\section*{Acknowledgments}
The authors thank M.\ Hirsch for discussions and suggestions.
This work has been supported by the DFG, project nr.\ PO-1337/2-1. F.T.\ has
also been supported by the DFG research training group GRK 1147.

\bibliography{brpv_fit}

\end{document}